\documentclass[sn-basic]{sn-jnl}

\usepackage[utf8]{inputenc}
\usepackage{amsmath}
\usepackage{mathtools}
\newcommand\numberthis{\addtocounter{equation}{1}\tag{\theequation}}

\jyear{2022}%

\theoremstyle{thmstyleone}%
\def\logR{\ensuremath{\log R^{\prime}_{\mathrm{HK}}}}

\theoremstyle{thmstyletwo}%

\theoremstyle{thmstylethree}%

\begin{document}

\title[The Colorado Ultraviolet Transit Experiment (CUTE) signal to noise calculator]{The Colorado Ultraviolet Transit Experiment (CUTE) signal to noise calculator}

\author*[1,2]{\fnm{A. G.} \sur{Sreejith}}\email{sreejith.aickara@oeaw.ac.at}
\author[1]{\fnm{Luca} \sur{Fossati}}
\author[1,3]{\fnm{P. E.} \sur{Cubillos}}
\author[2]{\fnm{S} \sur{Ambily}}
\author[2]{\fnm{Brian} \sur{Fleming}}
\author[2]{\fnm{Kevin} \sur{France}}

\affil*[1]{\orgdiv{Space Research Institute}, \orgname{Austrian Academy of Sciences}, \orgaddress{\street{Schmiedlstrasse 6}, \city{Graz}, \postcode{8042}, \country{Austria}}}

\affil[2]{\orgdiv{Laboratory for Atmospheric and Space Physics}, \orgname{University of Colorado}, \orgaddress{\city{Boulder}, \postcode{80303}, \state{CO}, \country{USA}}}

\affil[3]{\orgname{INAF – Osservatorio Astrofisico di Torino}, \orgaddress{\street{Via Osservatorio 20}, \city{Pino Torinese}, \postcode{10025}, \country{Italy}}}

\abstract{We present here the signal-to-noise (S/N) calculator developed for the Colorado Ultraviolet Transit Experiment (CUTE) mission. CUTE is a 6U CubeSat operating in the near-ultraviolet (NUV) observing exoplanetary transits to study their upper atmospheres. CUTE was launched into a low-Earth orbit in September 2021 and it is currently gathering scientific data. As part of the S/N calculator, we also present the error propagation for computing transit depth uncertainties starting from the S/N of the original spectroscopic observations. The CUTE S/N calculator is currently extensively used for target selection and scheduling. The modular construction of the CUTE S/N calculator enables its adaptation and can be used also for other missions and instruments.  
}
\keywords{Software, Exoplanet, Signal-to-noise, ultraviolet}

\maketitle

\section{Introduction}\label{sec:intro}
The detection of the first exoplanets has opened a new field in astrophysics \citep{mayor1995,Charbonneau2000} and the finding that numerous exoplanets orbit close to their host stars has fostered studies of planet atmospheric mass loss \cite[e.g.][]{vidal2003,lammer2003}. These have then led to the conclusion that atmospheric escape plays a pivotal role in the long-term evolution of planetary atmospheres and in sculpting the observed exoplanet population \cite[e.g.][]{jin2014,jin2018,owen2018,kubyshkina2020}.

The majority of the observational studies of exoplanet atmospheric escape have been performed at ultraviolet (UV) wavelengths by the Hubble Space Telescope (HST). These have led to the detection of both hydrogen and metals (Mg, Fe etc.) in the upper atmospheres, beyond the planetary Roche lobe, for a number of close-in planets \cite[e.g.][]{vidal2004,fossati2010,ehrenreich2015,sing2019,cubillos2020,garcia2021}. These observations have demonstrated that the upper atmospheres of close-in giant planets are hydrodynamically escaping as a result of the absorption, and consequent atmospheric heating, of the stellar high-energy (X-ray and extreme ultraviolet) emission \cite[e.g.][]{yelle2004,garcia2007,koskinen2014}. More recently, the He{\sc i} metastable near infrared triplet at $\approx$10830\,\AA, which is observable from the ground, has become an important tracer of atmospheric escape, but the strong dependence of the presence and strength of planetary He{\sc i} absorption on stellar type and He abundance \citep{oklopcic2019,fossati2022} leaves UV transmission spectroscopy as the main workhorse to study atmospheric mass loss.

HST is the only facility currently available for studying exoplanetary upper atmospheres and escape at UV wavelengths. However, the shared-use nature of HST limits the number of planets that can be observed and, more importantly, the number of transits that can be observed for each planet, preventing for example to study transit variability in association with stellar variability due to activity. The Colorado Ultraviolet Transit Experiment is a 6U CubeSat mission fully dedicated to observe exoplanetary transits at near-UV wavelengths to study upper atmospheres and escape \citep{fleming2018,france2023,egan2023,sreejith2022}. 

We present here the CUTE signal-to-noise (S/N) calculator, which is available both as a website\footnote{\tt https://cute-snr.iwf.oeaw.ac.at/} or as standalone downloadable software\footnote{\tt https://github.com/agsreejith/CUTE-SNR}. This tool is extensively being used by the CUTE science and instrument teams to guide target selection. In Section~\ref{sec:imp}, we present the implementation of the S/N calculator. In Section~\ref{sec:web}, we describe the web interface and its development, while the summary and the conclusions are presented in Section~\ref{sec:summary}.

\begin{figure}
\begin{center}
\includegraphics[width=0.95\textwidth]{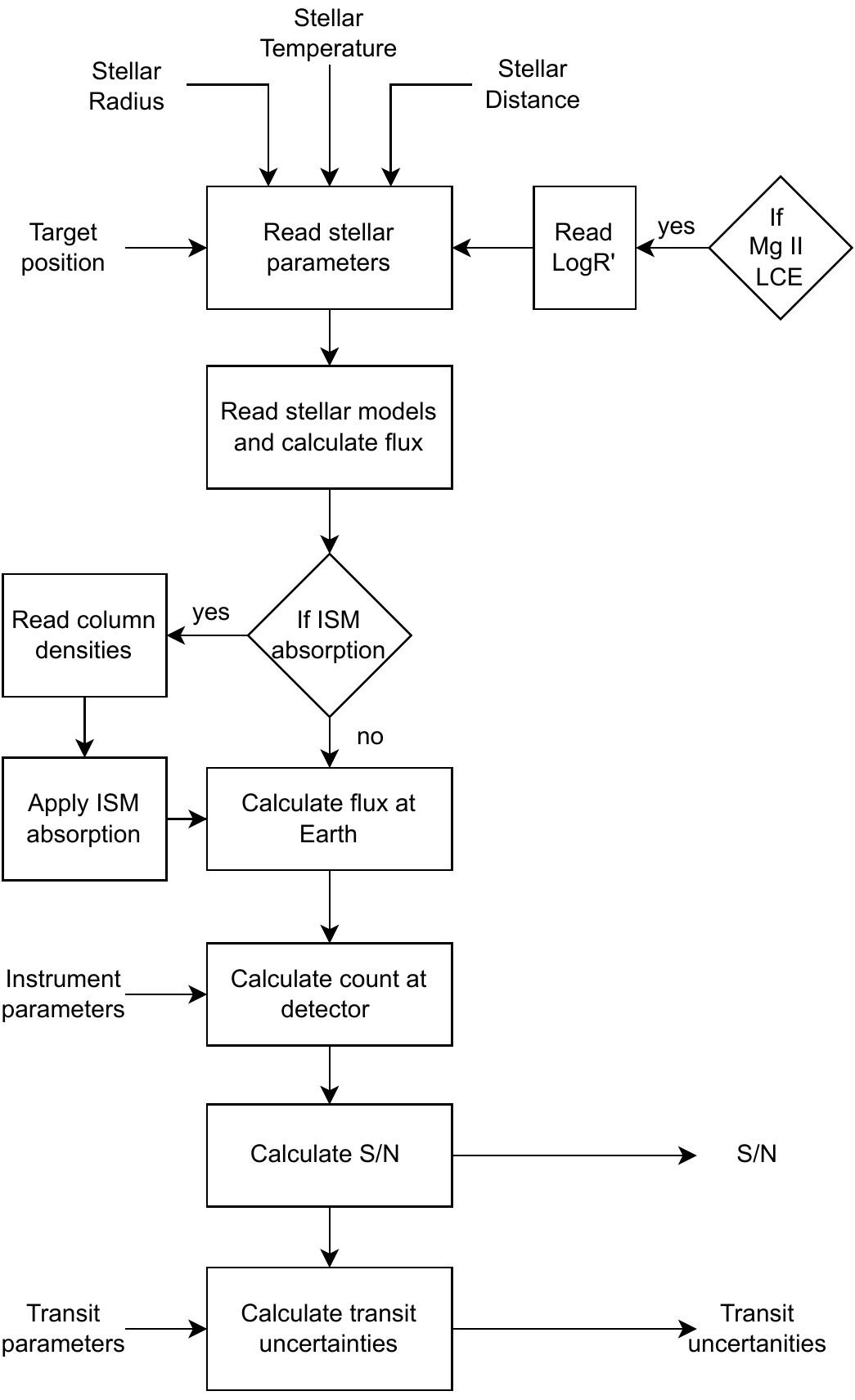}
\caption{Flowchart of the CUTE S/N calculator. In the top-right, LCE stands for ``line core emission''.} 
\label{fig:flowchart}
\end{center}
\end{figure}
%
\section{Implementation}
\label{sec:imp}
\subsection{General structure}

The CUTE S/N calculator is an offshoot of the CUTE data simulator \citep{sreejith2019}, which takes stellar models, stellar parameters, instrument specifications, and planetary transit information as input to generate simulated CUTE data that can be used to simulate transit lightcurves in the CUTE band, including their uncertainties. Figure~\ref{fig:flowchart} presents the overall working flowchart of the CUTE S/N calculator. The code reads the input parameters to generate a spectrum of the target star in the CUTE band by employing a library of synthetic photospheric stellar fluxes. The available library of stellar flux models has been generated employing the LLmodels stellar atmosphere code \citep{Shulyak2004}, which computes photospheric fluxes of stars assuming local thermodynamical equilibrium (LTE). This library covers stars raging between 3500\,K and 12000\,K, in steps of 100\,K below 6800\,K and in steps of 200\,K above it, with a wavelength sampling of 0.005\,\AA\ between 1500 and 9000\,\AA.

The CUTE spectrograph also covers the Mg{\sc ii}\,h\&k resonance lines, which in late-type stars present chromospheric line core emission (LCE) with a strength proportional to the stellar activity. The code automatically estimates the Mg{\sc ii}\,h\&k line core emission from the $\ensuremath{\log R^{\prime}_{\rm HK}}$ value and the stellar radius given in input \citep{fossati2017,sreejith2020}. The stellar flux at Earth is then derived by scaling for the distance to the star calculated from the parallax provided by the user. 

The CUTE S/N calculator also accounts for interstellar medium (ISM) absorption. In particular, we implemented both broadband extinction and ISM line absorption for Mg and Fe at the position of the NUV Mg{\sc i} (2852.127\,\AA), Mg{\sc ii} (2795.528\,\AA\ and 2802.705\,\AA), and Fe{\sc ii} (2599.395\,\AA) resonance lines \citep{sreejith2019}. Each ISM absorption feature is simulated as a single Voigt profile with a $b$-parameter (i.e. broadening) of 3\,km\,s$^{-1}$ \citep{Malamut2014} and assuming for no radial velocity shift between the stellar and ISM features as radial velocity shift has an negligible impact on transit measurements at the CUTE resolution \citep{sreejith2023}. The Mg{\sc i}, Mg{\sc ii}, and Fe{\sc ii} ISM column densities ($\log N_{\rm ion}$), which set the strength of the ISM absorption features, are required input from the user, who can use the algorthm described by \citep{sreejith2019} to estimate the ISM column densities for each individual species.

The stellar flux at Earth, optionally modified to account for the ISM absorption, is then converted from erg\,cm$^{-2}$\,s$^{-1}$\,\AA$^{-1}$ to photons\,cm$^{-2}$\,s$^{-1}$\,\AA$^{-1}$. The spectra are then trimmed to the relevant wavelength range (2478--3305\,\AA\ in the case of CUTE), multiplied by the effective area of the instrument, and convolved to the spectrograph's resolution, finally obtaining the CCD detector counts (counts\,\AA$^{-1}$\,s$^{-1}$). As a further step, the S/N calculator converts the counts\,\AA$^{-1}$\,s$^{-1}$ to counts\,pixel$^{-1}$\,s$^{-1}$ by taking into account the spectral resolution and the number of pixels per resolution element. Finally, the CCD counts are multiplied by the exposure time to obtain the expected signal from a particular source. The CCD counts are further converted from counts to photo-electrons using the gain provided by the user. The noise calculation is described in Section~\ref{sec:noise}, while here below we detail the input parameters requested by the code.

\subsection{Input Parameters}
\label{sec:ip}
\subsubsection{Stellar Parameters}
\noindent {\bf Temperature:} Temperature of the star in Kelvin. The stellar model to employ is selected on the basis of the given temperature and it sets the shape of the stellar spectral energy distribution (SED).

\noindent {\bf Radius:} Radius of the star in solar radii.

\noindent {\bf Coordinates:} Right ascension and declination of the star in degrees in J2000. This information is used to calculate the extinction and ISM column densities.

\noindent {\bf Distance:} Distance to the star as parallax in milliarcseconds (mas). This is used to scale the stellar SED at Earth.

\noindent {\bf Stellar activity index:} Stellar activity index (\logR) of the star. This parameter is used to set the strength of the Mg{\sc ii}\,h\&k line core emission.

\noindent {\bf Line core emission:} If this option is enabled, it adds the line core emission at the position of Mg{\sc ii}\,h\&k resonance lines assuming that the behaviour of the Mg{\sc ii}\,h\&k emission with stellar temperature is comparable to that of the Ca{\sc ii}\,H\&K emission.

\noindent {\bf Interstellar medium absorption:} The S/N calculator adds extinction following the method described in \cite{sreejith2019}. The user has the option to add ISM absorption at the position of any of the three Mg and Fe main resonance lines in the CUTE band: Mg{\sc i} (2852.127\,\AA), Mg{\sc ii} (2795.528\,\AA\ and 2802.705\,\AA), and Fe{\sc ii} (2599.395\,\AA). If ISM absorption is enabled, the user has to provide the column densities of the corresponding species and to estimate them users can either use the method described in \cite{sreejith2019} or employ the {\sc calc\_cd} code available with the CUTE S/N software distribution.
\subsubsection{Instrumental Parameters}
\noindent {\bf Spectral Resolution:} Spectral resolution of the spectrograph in angstroms.

\noindent {\bf CCD readout noise:} The CCD readout noise in electrons\,pixel$^{-1}$.

\noindent {\bf CCD dark noise:} The CCD dark noise in electrons\,pixel$^{-1}$\,s$^{-1}$.

\noindent {\bf CCD gain:} The CCD gain in photoelectrons\,ADU$^{-1}$.

\noindent {\bf Spectrum width:} Expected width in the cross-dispersion direction of the spectrum in pixels.

\noindent {\bf Exposure time:} Exposure time for each observation in seconds.

\noindent {\bf CCD read time:} The CCD read time for each observation in seconds.

In addition to these parameters, the calculator requires the effective area of the instrument and the wavelength to pixel solution. In the case of CUTE, these profiles are available with the software distribution and are those used also by the web interface.
\subsubsection{Transit Parameters}
The exoplanetary transit parameters required by the S/N calculator are the {\bf transit duration} of the planet in hours and the total {\bf number of transit observations}.
\subsubsection{Wavelength Parameters}
By default the S/N calculator computes the noise level in seven bands, as described in table~\ref{table:1}.  In addition to these bands, the user can specify a maximum of 20 wavelength bands of specific interest.

\begin{table}
\caption{Wavelength bands for CUTE S/N calculator.}     
\label{table:1}
\centering                                      
\begin{tabular}{l c }          
\hline\hline                        
Band & Range (in \AA)  \\    
\hline                                  
   Full Band & 2478.48--3304.86\, \\
   Lower Band & 2478.48--2788.84\,\\
   Mid Band & 2789.24--3059.57\,\\
   Upper Band & 3059.97--3304.86\, \\
   Mg{\sc ii} Band  & 2792.81--2804.72\, \\
   Mg{\sc i} Band & 2849.97--2853.54\, \\
   Fe{\sc ii} Band & 2582.81--2586.78\, \\
 \hline                                            
\end{tabular}
\end{table}

%
\subsection{Noise Propagation}
\label{sec:noise}

The S/N calculator considers three main sources of noise. 

\noindent {\bf Read noise} of the CCD detector. In most of the modern cooled scientific CCDs, the readout noise sets a limit on the detector's performance and can be reduced at the expense of increasing readout time. 

\noindent {\bf Dark noise} in the detector due to thermal CCD electrons. Dark noise is reduced by operating the CCD at low temperatures inhibiting the creation of thermal electrons.

\noindent {\bf Photon noise} that is assumed to follow a Poissonian distribution, and thus computed as the square root of the number of incoming photons. 

Considering these noise components the $S/N$ ratio of the stellar spectrum extracted from the CCD, $\frac{S}{N}\big{\vert}_{F}$ is
\begin{equation}
\frac{S}{N}\Big{\vert}_{F}  = \frac{F}{\sigma_F}.\label{eq1}
\end{equation}
where $F$ is the integrated CCD counts in photo-electrons in the wavelength range of interest and $\sigma_F$ is the related uncertainty. In particular,
\begin{equation}
    F =  \sum{\Delta \lambda F_\lambda}
\end{equation}
and
\begin{equation}
    \sigma_F^2 =   \sum{(\Delta \lambda\,\,\sigma_{F_{\lambda}})^2}\,,
\end{equation}
where $\Delta \lambda$ is the considered wavelength range, $F_\lambda$ is the expected counts value (in photo-electrons) as a function of wavelength and $\sigma_{F_{\lambda}}$ is the respective uncertainty that is computed as
\begin{equation}
\label{eq:noise1}
   \sigma_{F_{\lambda}} = \sqrt{{\rm p\_noise_{\lambda}}^2 + {\rm s\_width}\,({\rm r\_noise}^2+({\rm d\_noise_{\lambda}} \times {\rm exptime} \times G))}\,.
\end{equation}
In Equation~(\ref{eq:noise1}), p\_noise$_{\lambda}$ is the photon noise associated with the signal at each wavelength ${\lambda}$, r\_noise is the read noise of the CCD in ADU, d\_noise$_{\lambda}$ is the dark noise of the CCD in electron\,s$^{-1}$\,pixel$^{-1}$, exptime is the exposure time of the observation, $G$ is the CCD gain, and s\_width specifies the cross-dispersion width of the spectrum in pixels. All these parameters are user inputs except for the photon noise which is calculated as $\sqrt{F_\lambda}$.
\subsection{Transit uncertainties}
\label{sec:transit}
The S/N calculator then combines the available information to compute the final uncertainty on the transit depth, which the instrument is expected to be capable to measure following one transit observation or by combining multiple transits, as follows. The S/N ratio of the transit depth measurement is
\begin{equation}
    \frac{S}{N}\Bigg{\vert}_{d} = \frac{d}{\sigma_d}\,,
\end{equation}
where $d$ is the transit depth and $\sigma_d$ its uncertainty. The former can be written as
\begin{equation}
\label{eq:depth}
    d = 1-\frac{F_{\rm in}}{F_{\rm out}} = \left(\frac{R_{\rm p}}{R_{\rm s}}\right)^2\,,
\end{equation}
where $F_{\rm in}$ is the in-transit flux, $F_{\rm out}$ is the out of transit flux, $R_{\rm p}$ is the planetary radius, and $R_{\rm s}$ is the stellar radius. Therefore, the uncertainty on the transit depth, ignoring the uncertainty on the stellar radius, can be written as
\begin{equation}
    \sigma_d = \frac{\partial d}{\partial R_{\rm p}}\sigma_{R_{\rm p}} = \frac{2R_{\rm p}}{R_{\rm s}^{2}} \sigma_{R_{\rm p}} = \frac{2d}{R_{\rm p}}\sigma_{R_{\rm p}}\,,
\end{equation}
where $\sigma_{R_{\rm p}}$ is the uncertainty on the planetary radius, which then becomes
\begin{equation}
    \sigma_{R_{\rm p}} = \frac{R_{\rm p}}{2d}\sigma_d = \frac{R_{\rm p}}{2\frac{S}{N}\big\vert_d}\,.
\end{equation}

 By combining Equation~(\ref{eq:depth}) and Equation~(\ref{eq1}), one obtains
\begin{align*}
    \sigma_d &= \sqrt{\left(\frac{\partial d}{\partial F_{\rm out}}\right)^2
    \sigma_{F_{\rm out}}^2+
    \left(\frac{\partial d}{\partial F_{\rm in}}\right)^2 \sigma_{F_{\rm in}}^2} \\
    & = \sqrt{\left(\frac{F_{\rm in}}{F_{\rm out}^2}\right)^2 \sigma_{F_{\rm out}}^2+
    \left(\frac{1}{F_{\rm out}}\right)^2 \sigma_{F_{\rm in}}^2}\,,  \numberthis
\end{align*}
where $\sigma_{F_{\rm in}}$ and $\sigma_{F_{\rm out}}$ are the uncertainties on the in- and out of transit fluxes, respectively. Then, assuming $F_{\rm in}$\,$\approx$\,$F_{\rm out}$\,$\equiv$\,$F$ and $\sigma_{F_{\rm in}}$\,$\approx$\,$\sigma_{F_{\rm out}}$\,$\equiv$\,$\sigma_F$, one obtains
\begin{equation}
    \sigma_d = \sqrt{\left(\frac{F}{F^2}\right)^2 \sigma_{F}^2+
    \left(\frac{1}{F}\right)^2 \sigma_{F}^2} 
    = \sqrt{2\left(\frac{1}{F}\right)^2 \sigma_{F}^2}
    = \sqrt{2}\frac{\sigma_{F}}{F}
    = \frac{\sqrt{2}}{\frac{S}{N}\big{\vert}_{F}}\,. \label{eqn}
\end{equation}
Therefore,
\begin{equation}
     \frac{S}{N}\Big{\vert}_{d} = \frac{d}{\sqrt{2}\frac{\sigma_F}{F}} 
     = \frac{d}{\sqrt{2}} \frac{S}{N}\Big{\vert}_{F}\,, \label{eqn}
\end{equation}
which implies that the uncertainty on the planetary radius as a function of the S/N of the observed spectra is
\begin{equation}
    \sigma_{R_{\rm p}} = \frac{R_{\rm p}}{2\frac{d}{\sqrt{2}}\frac{S}{N}\big{\vert}_{F}}\,. \label{eqn}
\end{equation}
%
%
\section{Web Interface}
\label{sec:web}
\begin{figure}[h]
\begin{center}
\includegraphics[width=\textwidth]{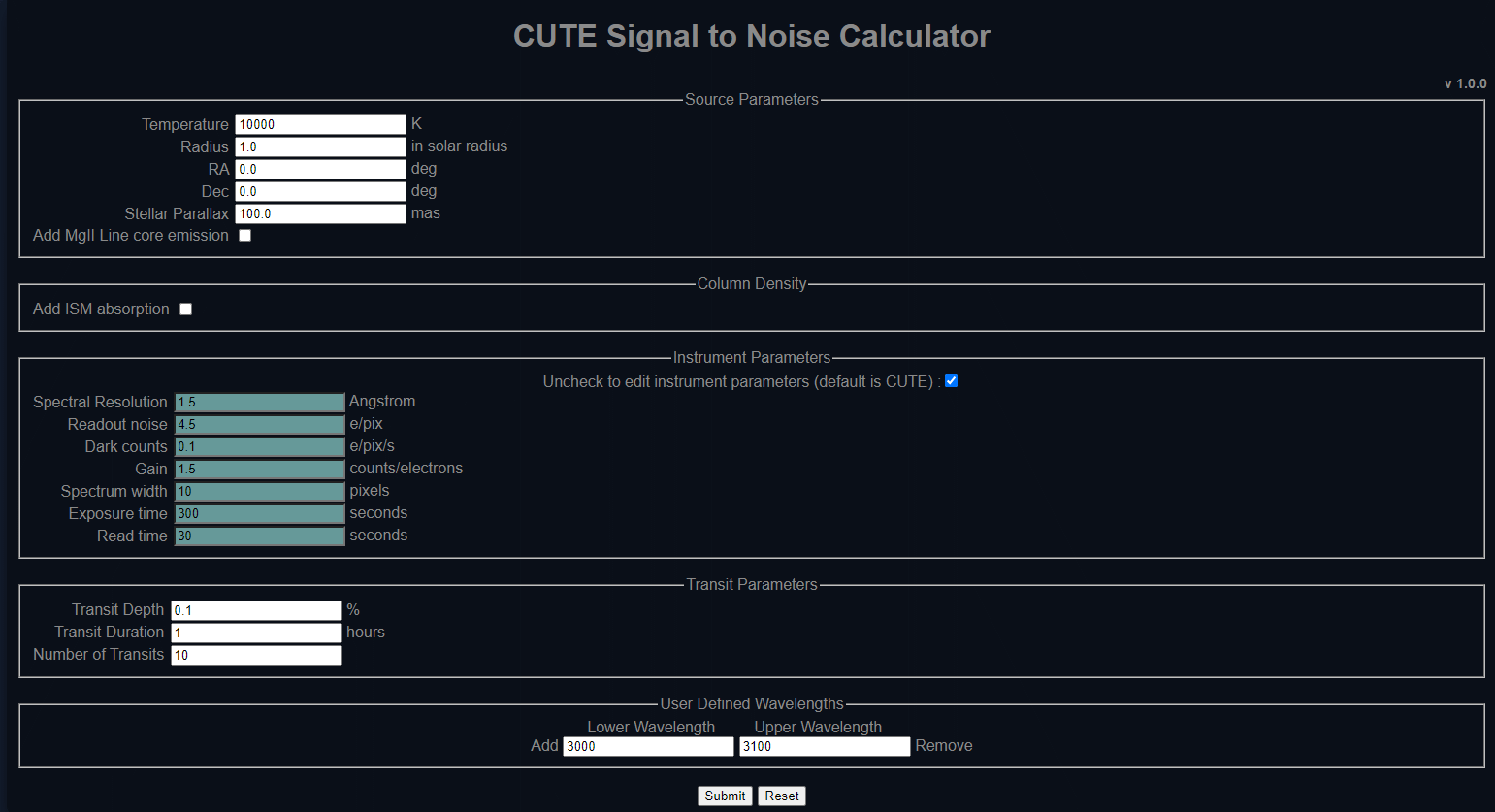}
\caption{Input web page of the CUTE S/N calculator.} 
\label{fig:web1}
\end{center}
\end{figure}

The web application of the CUTE S/N calculator has been developed on Python Flask web framework, which makes use of the Jinja template engine and the Werkzeug (WSGI) toolkit. Figure~\ref{fig:web1} shows the main web page of the S/N calculator. The web application accepts input values from the user, validates the inputs, and run the S/N calculator, finally providing the user with the results. The validation of the input values is carried out by the server and users are notified in case of problems. The server reads the user input form and generates the necessary parameters for the S/N calculations. 

\begin{figure}[h]
\begin{center}
\includegraphics[width=\textwidth]{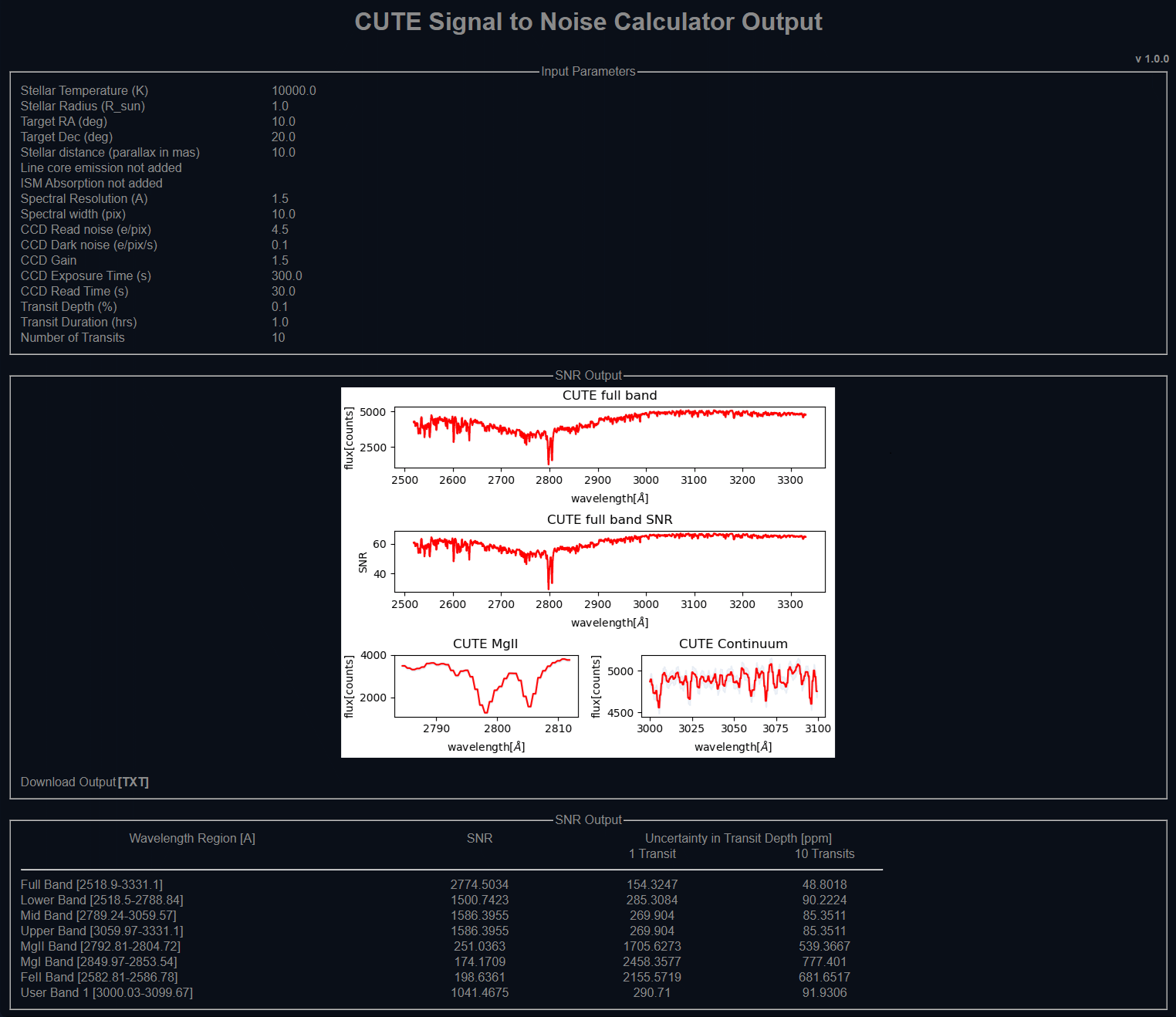}
\caption{Example output web page of the CUTE S/N calculator.} 
\label{fig:web2}
\end{center}
\end{figure}

\begin{figure}[h]
\begin{center}
\includegraphics[width=\textwidth]{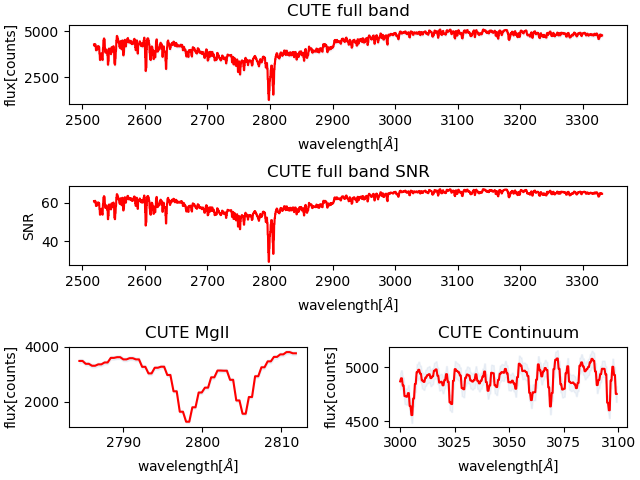}
\caption{Blow up of the plot shown in Figure~\ref{fig:web2}.} 
\label{fig:web3}
\end{center}
\end{figure}

The output web page, shown in Figure~\ref{fig:web2}, displays an example result of a S/N calculation, which contains also a quick look image that we show in Figure~\ref{fig:web3}. The basic output given on the web page is that of the S/N ratio and relative uncertainties on the transit depth as a function of wavelength. Furthermore, the S/N calculator enables the user to download a file detailing all outputs of the simulation for further analysis.
%
\section{Summary and Conclusion}
\label{sec:summary}
We presented the CUTE S/N calculator and its implementation as a web form. The modular approach of the calculator enables one to easily modify it to the needs of future missions and instruments carrying out transmission spectroscopy. The development of a web version also enables one to use the calculator without the need to install extra software or computational resources. 

In the future, we plan to introduce additional features that would make the S/N calculator more general and easy to use, the most important being
\begin{itemize}
\item Option to upload a stellar model file;
\item Option to upload both effective area and wavelength solution files;
\item Option to consider ISM absorption at other wavelengths;
\item Option to derive stellar parameters from the target name, for example through querying the latest GAIA catalog.
\item Evaluating the effect of different scatter light on S/N calculations.

\end{itemize}
\section*{Acknowledgements}
We thank Wolfgang Voller for his help in setting up the website and the server.
L.F. acknowledge financial support from the Austrian Forschungsf\"orderungsgesellschaft FFG project CARNIVALS P885348. This project was partly funded by the Austrian Science Fund (FWF) [J 4596-N]. P.C. acknowledges support and funding by the Austrian Science Fund (FWF) Erwin Schroedinger Fellowship program J4595-N. We thank the anonymous referee for their comments that helped to significantly improve the paper.

\bibliography{ref.bib}



\end{document}